# Psychological constraints on string-based methods for pattern discovery in polyphonic corpora

David R. W. Sears[1], Gerhard Widmer[2]
[1] College of Visual & Performing Arts, Texas Tech University, Lubbock, TX, USA
[2] Department of Computational Perception, Johannes Kepler University, Linz, AUSTRIA

**Background**

Researchers often divide symbolic music corpora into contiguous sequences of $n$ events (called $n$-grams) for the purposes of pattern discovery, key finding, classification, and prediction. Several studies have reported improved task performance when using psychologically-motivated weighting functions, which adjust the count to privilege $n$-grams featuring more salient or memorable events (e.g., Krumhansl, 1990). However, these functions have yet to appear in algorithms that attempt to discover the most recurrent chord progressions in complex polyphonic corpora.

**Aims**

This study examines whether psychologically-motivated weighting functions can improve harmonic pattern discovery algorithms. Models using various $n$-gram selection methods, weighting functions, and ranking algorithms attempt to discover the most conventional closing progression in the common-practice period, ii6-"I64"-V7-I, with the progression's mean reciprocal rank serving as an evaluation metric for model comparison.

**Methods**

The corpus features 275 pieces of symbolic Western classical music that were time-aligned to audio recordings of expressive performances. To derive chord progressions, we performed a *full expansion* of the symbolic encoding, which duplicates overlapping note events at every unique onset time (Conklin, 2002). We then applied the *voice-leading type* representation (Quinn, 2010), which produces an optimally reduced and key-invariant chord typology that models every possible combination of note events in the corpus.

The pattern discovery pipeline consists of the following parameters:
1) *Skip-grams* – Include $n$-grams whose constituent events occur either within a fixed number of skips (*fixed*; up to 0, 1, 2, 3, or 4 skips), or within a specified temporal boundary (*variable*; up to 0.5, 1, 1.5, or 2 s between event onsets).
2) *Weighted counts* – Weight the count for each $n$-gram on the real-unit interval [0,1], assigning higher weights to $n$-grams with temporally proximal event onsets (*proximity*), periodic inter-onset intervals (*periodicity*), or inter-onset intervals close to the periodicities at which listeners tend to tap (*resonance*).
3) *Ranking* – Rank each distinct $n$-gram type in the distribution using a family of information-theoretic attraction measures from corpus linguistics: *pairwise mutual information* (*PMI*), *directed PMI*, *local PMI*, and *piece-weighted PMI*.

**Results**

The cadential progression, ii6-"I64"-V7-I, obtained the highest rank for (1) skip-grams including up to two or three skips, and which were (2) weighted according to the periodicity of their constituent inter-onset intervals, and (3) ranked according to piece-weighted PMI.

**Conclusions**

This study demonstrates that applying psychological constraints to pattern discovery algorithms improves task performance. These methods also reveal the temporal interval over which recurrent progressions appear with significant frequency in polyphonic corpora.

**Acknowledgements**

This project has received funding from the European Research Council (ERC) under the European Union's Horizon 2020 research and innovation programme (grant agreement n° 670035).